\documentclass[showpacs,preprintnumbers,amsmath,amssymb]{revtex4}
\usepackage{epsfig}

\newcommand{\be}{\begin{equation}}
\newcommand{\ee}{\end{equation}}
\newcommand{\bea}{\begin{eqnarray}}
\newcommand{\eea}{\end{eqnarray}}

\begin{document}
\vspace*{4cm}
\title{Exploring the possibility of detecting dark energy in a terrestrial experiment using atom interferometry}

\author{Martin L. Perl }

\address{SLAC National Accelerator Laboratory, Stanford University, 2575 Sand Hill Road, Menlo Park, California 94025, USA}

\author{ Holger Mueller}

\address{Department of Physics, University of California, Berkeley, California 94720, USA}

\begin{abstract}The majority of astronomers and physicists accept the reality of dark energy but also believe it can only be studied indirectly through observation of the motions of galaxies. This paper opens the experimental question of whether it is possible to directly detect dark energy on earth using atom interferometry through a force hypothetically caused by a gradient in the dark energy density. Our proposed experimental design is outlined. The possibility of detecting other weak fields is briefly discussed.
\end{abstract}
\maketitle
\section{Introduction}

This is a revised version of a paper presented at the Windows on the Universe, XXI Rencontres de Blois, 21-26 June, 2009. This paper is an exploration of the possibility of detecting dark energy in a terrestrial experiment using atom interfereometry. Since then we have begun designing preliminary tests of this concept.

In Secs. 1 and 2 we present an Introduction and summarize conventional beliefs about the nature of dark energy. In Sec. 3 we present an illuminating comparison of dark energy density with the energy density of a weak electric field. In Sec. 4 we review how the precision of present measurements of the earth's gravitational force field set an upper limit on a possible dark energy force field. In Sec. 5 we present our consideration of how much we can improve that precision. In Secs. 6, 7 and 8 we present our assumptions about the properties of dark energy that would make the experiment feasible, a brief description of our experimental method, and remarks on other unknown weak forces. We include an Appendix on other suggestions for experimental searches for dark energy.

For pedagogical purposes we list the following quantities and parameters:

\begin{itemize}
\item {Critical energy density = $\rho_{crit}$ = $9\times10^{-10}$  J/m$^{3}$}. The critical energy density is the density of total mass-energy required for the universe to be flat.
\item{Dark energy density = $\rho_{dark energy}$ = $0.70\times\rho_{crit}$ = $6.3\times10^{-10}$  J/m$^{3}$}
\item{$\hbar$ = $1.054\times 10^{-34}$ Js}
\item{Newton's constant = G = $6.67\times10^{-11}$ m$^{3}$ kg$^{-1}$ s$^{-2}$}
\end{itemize}

We use MKS units in this paper because it is about laboratory experiments.

\section{Conventional beliefs about the nature of dark energy }

Present conventional beliefs about dark energy density are that it is approximately uniformly distributed in space and that its magnitude is given by $6.3\times10^{-10}$  J/m$^{3}$. The usual assumption is that every cubic meter of space contains the same dark energy density so that as the visible universe expands there is more total dark energy \cite{Peebles}. One of us  (MLP) trained many years ago in thermodynamics at Columbia University, is bothered by this apparent breach of the principle of the conservation of energy.

$\rho_{dark energy}$ = $6.3\times10^{-10}$  J/m$^{3}$ initially strikes us as a very small energy density but as shown in the next section we experiment with smaller electric field energy densities in the laboratory.

\section{Comparison of dark energy density with the energy density of a weak electric field}

Consider a weak electric field, for example $E$ = 1 volt/m. Using

\begin{equation}
\rho_{electric field} = \epsilon_{0}E^{2}/2
\end{equation}

\begin{equation}
\rho_{electric field}=4.4\times10^{-12} \mbox{  J/m$^3$}
\end{equation}

\noindent Hence the energy density of this electric field is 100 times smaller than the dark energy density, $\rho_{dark energy}$ = $6.3\times10^{-10}$  J/m$^{3}$, yet this weak electric field is easily detected and measured. Thus we work with fields whose energy densities are much less than $\rho_{dark energy}$. This realization started one of us(MLP) thinking about the possibility of direct detection of dark energy

Of course, it is easy to sense tiny electromagnetic fields using electronic devices such as field effect transistors or superconducting quantum interference devices (SQUIDs).

On the other hand there are obviously severe experimental problems in detecting the dark energy density.
\begin{itemize}
\item{Unlike an electric field in the laboratory, we cannot turn dark energy on and off.}
\item{We do not know if there is a zero dark energy field to use as an experimental reference. In the fixed value, cosmological constant explanation of dark energy, $\rho_{dark energy}$ has the same value in all space.}
\item{Even if the dark energy density should have a gradient, what force does it exert on a material object of mass $M$? From general relativity, we know that the mass dark energy density equivalent of $\rho_{dark energy}$ will exert a gravitational force on $M$, but is there any other and larger force?}
\end{itemize}

\section{The terrestrial gravitational force field and a possible dark energy force}

We originally considered comparing dark energy density with gravitational energy density and tried to do so in the first version of this paper, but we then realized that in general relativity there is no standard definition of local gravitational energy density. For example in one space-time frame gravitational force may exist at some point, but one can transform to an accelerated frame where the gravitational force is zero and the local gravitational energy density is zero. Therefore we use forces rather than energy density for comparison.

In atom interferometry the phase change of atoms depends upon the integral of the potential difference between two separate trajectories of the atom in space. Of course at present we know nothing about whether or not dark energy exerts such a force. Indeed investigating this question is one of the purposes of our proposed experiment. In analogy we designate this force as $g_{dark energy}$ in units of force per unit mass.

Three comments on $g$ and $g_{dark energy}$.

\begin{enumerate}
\item{The gravitational force per unit mass on earth is of course $g$=9.8 m/s$^2$.}
\item{Atom interferometry studies have reached a sensitivity of much better than $10^{-9}$ $g$ in measurements of the gravitational acceleration \cite{Chung} and found no anomaly. Even though a definite analysis for this has not be performed, it is probably safe to say that there is no evidence for $g_{darkenergy}$ at this level.}
\item{ Therefore $g_{darkenergy}\leq10^{-8}$ m/s$^2$} using our assumptions about the properties of dark energy enumerated in Sec. 6 .
\end{enumerate}

\section{Preliminary considerations on how well we can null out $g$.}

The equivalence principle prevents us from measuring a uniform density of dark matter (or dark energy) in a local experiment on Earth, if it interacts gravitationally only. However, detection will be possible if the density has spatial fluctuations on terrestrial length scales (which cannot be ruled out from existing observations). In addition the interaction of dark energy with matter could depend on the nature of the matter, such as composition, spin, and charge. \emph{Based on preliminary considerations we believe we can null out $g$ to a precision perhaps as small as $10^{-17}$. This sets the smallest $g_{dark energy}$ that we can investigate at $10^{-16}$ $m/s^2$.}

\section{Author's assumptions about the properties of dark energy that make the experiment feasible}

We assume:

\begin{itemize}
\item {A dark energy force, $F_{darkenergy}$, exists other than the gravitational force equivalent of $\rho_{darkenergy}$.}
\item {$F_{darkenergy}$ is sufficiently local and thus  $\rho_{darkenergy}$ is sufficiently non-uniform so that $F_{darkenergy}$ varies over a length of the order of a meter.}
\item {$F_{darkenergy}$ acts on atoms leading to a potential energy $V_{darkenergy}$.}
\item {The ratio $g_{darkenergy}/g$ is large enough for $g_{darkenergy}$ to be detected in this experiment by nulling signals from $g$.}
\end{itemize}

\section{Brief description of our experimental method}

The search for {$F_{darkenergy}$ requires the nulling of all the known forces that can change the atomic phase. The effects of electric and magnetic forces are nulled by shielding and by using atoms in quantum states which are not sensitive to the linear Zeeman and Stark effects.

Figure 1 shows the apparatus schematically, two identical atom interferometers are used with the solid lines representing atom beams and the dash lines representing signal flow.
In interferometer 1 for example, an atom beam from source O is split at so that each atom quantum mechanically follows the two paths ABC and ADC. At D the two states arrive with relative phases, $\phi_{ABC}$ and $\phi_{ADC}$. The interference produces a signal $T_{1}$ proportional to the phase difference $\phi_{ABC}$-$\phi_{ADC}$. $T_{1}$ depends upon the potentials acting on the atoms in the space ABCD.  Considering just the earth's gravitational force $\vec{g}$, $T_{1}$ is proportional to the change in gravitational potential between the upper path ABC and the lower path ADC , and thus proportional to the gravitational acceleration g..

\begin{figure}
\begin{center}
\epsfig{file=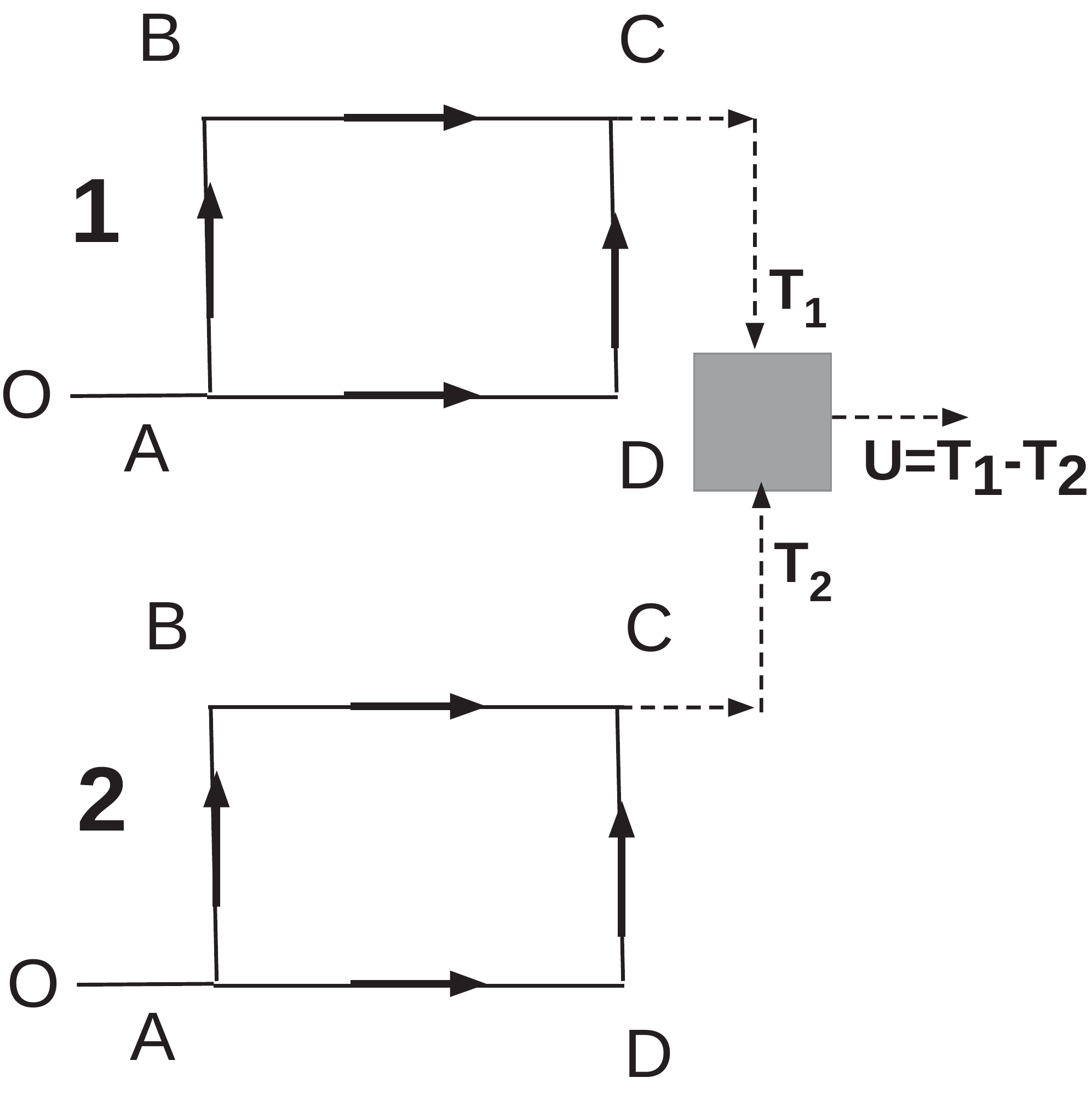,height=3in}
\caption{Schematic of proposed apparatus using two atom interferometers. See text for details.}
\end{center}
\end{figure}

$U$ = $T_{1}$-$T_{2}$ is given by the difference between the accelerations of free fall at the locations of the first and second interferometers. If we assume $\vec{g}$ to be nearly constant at the earth's surface, $U$ = 0 for contributions from $\vec{g}$, except for small corrections. Thus signals from the gravitational force are nulled by this interferometer design.

One realization of this design is a pair of fountain interferometers as described by one of us (HM) and his coworkers \cite{Chung}. Even in a single interferometer, suppression of the signal due to $g$ to the $10^{-10}\,g$ level has already been demonstrated by subtracting a Newtonian model of tidal variations caused by the Moon, the Sun, and the planets. Using the pair of interferometers described above, we expect to be able to cancel the effects of gravity by a factor of $10^{-17}$.

Figure 2 shows schematically how a non-uniform dark energy density can be detected. The vertical height of the apparatus will be on the order of one meter. Therefore, for detection, the non-uniformity must exist over this size range. However for length scales of the non-uniformity different from that range, a suppressed sensitivity can still be attained that scales with the ratio of the ranges.

\begin{figure}
\begin{center}
\epsfig{file=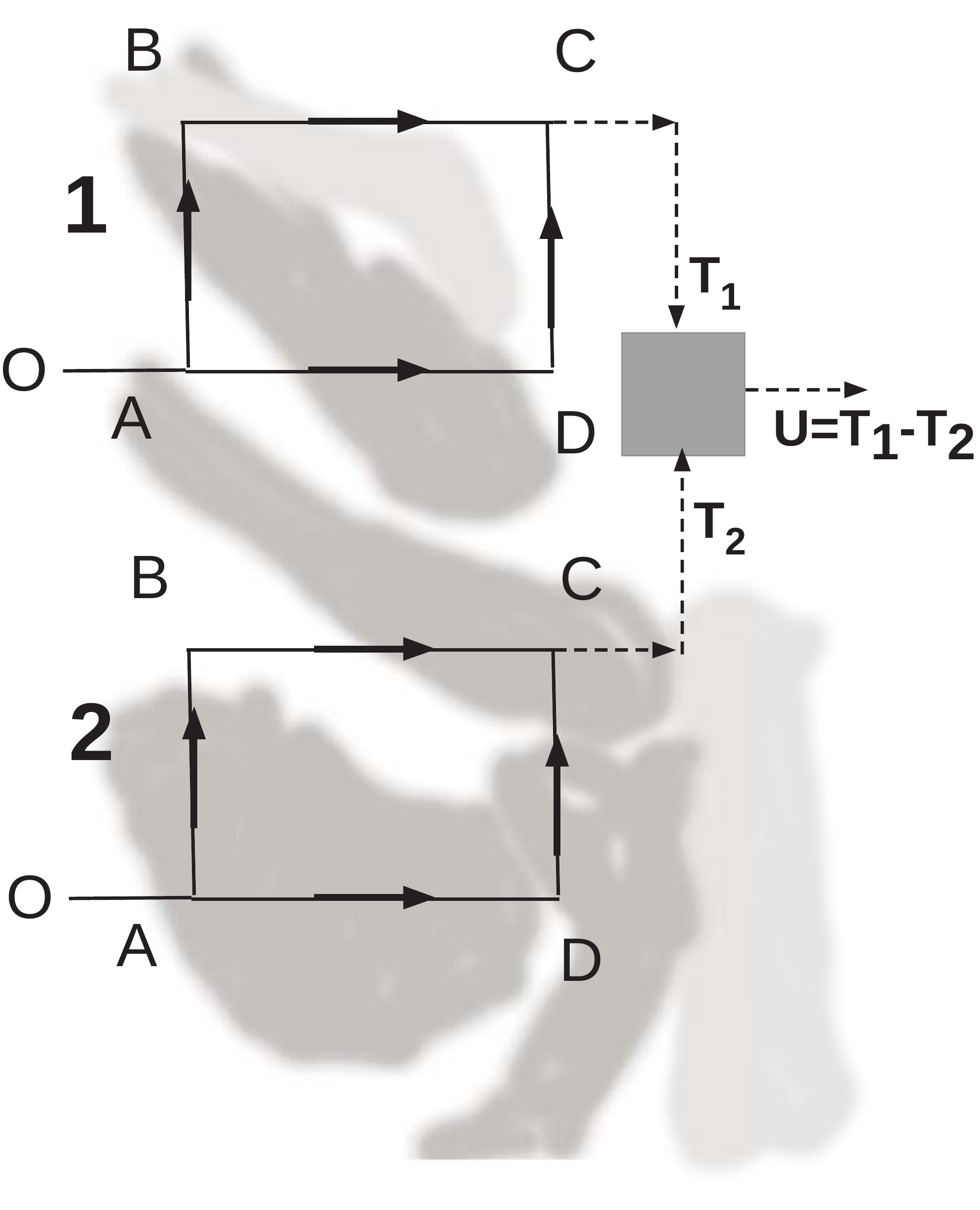,height=3in}
\caption{Schematic of how a non-uniform dark energy density can be detected. The difference in dark energy density between the shaded and unshaded areas may be a fraction of $\rho_{dark energy}$, as small as a few percent.}
\end{center}
\end{figure}

\section{Other unknown weak forces and remarks}

The foregoing discussion applies to other unknown weak forces. Hence this experiment is  a general probe into what fields or forces that may exist in space but have until now been beneath our view. But to our knowledge the experiment has nothing to do with the ongoing, vexing question of the relation of zero-point vacuum energy to dark energy.

\section*{Acknowledgments}
We are grateful to Professor Tran Van Thanh for the opportunity to present this paper at the XXI Rencontres de Blois and for discussions with many knowledgeable physicists at the conference.

\section*{Appendix}

 In this Appendix  section we briefly describe other terrestrial experimental searches for dark energy.

\textbf{Search for a high frequency cutoff in electromagnetic zero-point energy}
\newline Beck and McKay \cite{Beck} proposed that dark energy comes from a lower frequency part of the zero point energy of the electromagnetic field. They conjecture that this will be indicated by a decrease in the noise spectrum in superconductors for frequencies above $10^{12}$ Hz and propose using Josephson junction noise for the test. At present there is no definitive data on this noise in this frequency range. The theoretical underpinnings of this conjecture have been criticized by Jetzer and Straumann \cite{Jetzer}and by Doran and Jaeckel \cite{Doran}.

\textbf{Probing for non-astronomical evidence for dark energy using small distance gravitational force measurements}
\newline In the last decade there has been substantial interest in testing Newton's law of gravitation at distances less than a millimeter \cite{Smullin,Long,Kapner}. One motivation is to test the hypothesis that the existence of extra dimensions will change the behavior of the gravitational force at small distance. Another motivation is that the Newtonian behavior of the gravitational force at small distance might be affected by the so-called dark energy length $L_{darkenergy}$ taken to be about  84 microns. It is conventional to seek deviations from the Newtonian inverse square law using the following elaboration of the gravitational potential $V_{g}(r)$.

\begin{equation}
V_{g}(r) = -G(m_1m_2/r)\times[ 1+\alpha \mbox { exp}(-r/\lambda]
\end{equation}

The experimenters have searched for a non-zero value for $\alpha$ and a finite value for $\lambda$. No deviations from the inverse square law have been found.

\end{document}